\documentclass{pasj00}
\begin{document}
\SetRunningHead{H. Yamaguchi et al.}{Suzaku Observation of SN~1006}
\Received{2007/05/06}
\Accepted{2007/06/12}
\title{X-Ray Spectroscopy of SN~1006 with Suzaku}

\author{
Hiroya \textsc{Yamaguchi},\altaffilmark{1}
Katsuji \textsc{Koyama},\altaffilmark{1}
Satoru \textsc{Katsuda},\altaffilmark{2}
Hiroshi \textsc{Nakajima},\altaffilmark{2} 
John~P. \textsc{Hughes},\altaffilmark{3} \\
Aya \textsc{Bamba},\altaffilmark{4} 
Junko~S. \textsc{Hiraga},\altaffilmark{5} 
Koji \textsc{Mori},\altaffilmark{6} 
Masanobu \textsc{Ozaki},\altaffilmark{4} and 
Takeshi Go \textsc{Tsuru}\altaffilmark{1}
}

\altaffiltext{1}
{Department of Physics, Kyoto University,
Kitashirakawa-oiwake-cho, Sakyo-ku, Kyoto 606-8502}
\email{hiroya@cr.scphys.kyoto-u.ac.jp}
\altaffiltext{2}
{Department of Earth and Space Science, Osaka University, 
1-1 Machikaneyama, Toyonaka, Osaka 560-0043}
\altaffiltext{3}
{Department of Physics and Astronomy, Rutgers University, 136 
Frelinghuysen Road, Piscataway, NJ 08854-8019, U.S.A.}
\altaffiltext{4}
{Institute of Space and Astronautical Science, JAXA, 
3-1-1 Yoshinodai, Sagamihara, Kanagawa 229-8510}
\altaffiltext{5}
{RIKEN (The Institute of Physical and Chemical Research),
2-1 Hirosawa, Wako, Saitama 351-0198}
\altaffiltext{6}
{Department of Applied Physics, 
University of Miyazaki, 1-1 Gakuen Kibana-dai Nishi Miyazaki, 889-2192}


\KeyWords{Supernova Remnants:~individual~(SN~1006) 
--- X-Rays:~spectra

} 

\maketitle

\begin{abstract}
We report on observations of SN~1006 with Suzaku. 
We firmly detected K-shell emission from Fe, for the first time,
and found that the Fe ionization state is quite low.
The broad-band spectrum extracted from southeast of the remnant was 
well-fitted with a model consisting of three optically thin 
thermal non-equilibrium ionization plasmas and a power-law component. 
Two of the thermal models are highly overabundant in heavy elements, and
hence are likely due to ejecta. These components have different ionization 
parameters, $n_et \sim 1.4\times 10^{10}$~cm$^{-3}$~s and 
$n_et \sim 7.7\times 10^8$~cm$^{-3}$~s; it is the later one that
produces Fe-K emission.
This suggests that Fe has been heated by reverse shock 
more recently than the other elements, consistent with a picture where the
ejecta are stratified  by composition with Fe in the interior.
On the other hand, the third thermal component is 
assumed to be solar abundance, and we associate it 
with emission from the interstellar medium (ISM). 
The electron temperature ($kT_e \sim 0.5$~keV) is lower than 
that expected from the shock velocity, which suggests 
a lack of collisionless electron heating at the forward shock.
The extremely low ionization parameter and extreme non-equilibrium state 
are due to the low density of the ambient medium. 

\end{abstract}

\section{Introduction}
\label{sec:introduction}

\begin{figure}[tbh]
  \begin{center}
    \FigureFile(80mm,80mm){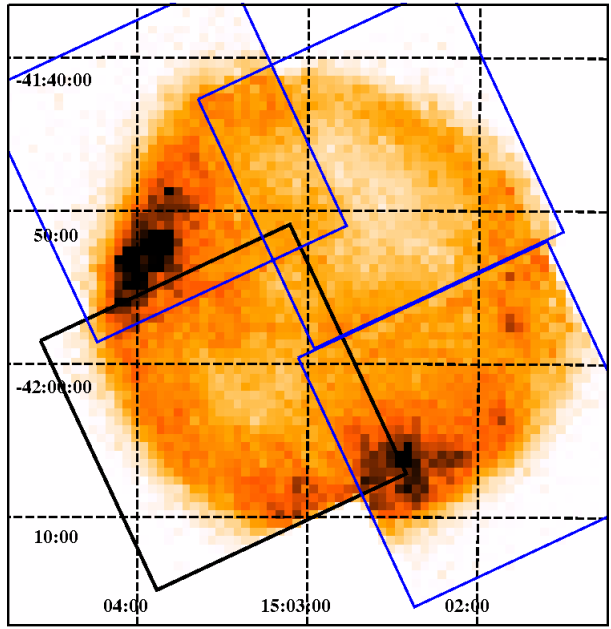}
  \end{center}
  \caption{XIS-BI mosaic image of SN~1006 
    in the 0.55--0.6 keV band. 
    Exposure and vignetting effects are corrected.
    The coordinates (R.A. and Dec) are referred to epoch J2000.0.
    The black square shows the FOV of the XIS on the SE region of SN~1006 
    where this paper is concentrated. 
    The blue squares show the FOVs on the other three quadrants 
    (NE, NW, and SW).}
  \label{fig:o_image}
\end{figure}

Based on the historical record, SN~1006 is widely regarded to be 
one of the Galactic type~Ia supernova remnants (SNRs) 
similar to Tycho's SNR (Schaefer 1996). 
The current supernova (SN) models predict that Fe production 
in type Ia SNe is far larger than that of core-collapse SNe 
(e.g., Nomoto et al.~1984; Iwamoto et al.~1999). 
Therefore, measuring the Fe abundance is one of the essential 
clues to classify the SN type of any specific remnant. 
The K-shell X-ray emission lines from ionized Fe offer 
the most direct information for  abundance determination.
In fact, strong Fe-K$\alpha$ lines had been observed 
from Tycho's SNR (e.g., Hwang et al.~1998; Decourchelle et al.~2001). 

In SN~1006, Vink et al.~(2000) suggested the possible presence of an
Fe-K emission line at 6.3$\pm$0.2~keV from the BeppoSAX MECS spectrum. 
However, the detection was not clear, 
because the energy resolution of the MECS (a gas scintillation counter) 
was relatively poor (FWHM$\sim$500~eV @6~keV). 
Other recent X-ray missions, Chandra and XMM-Newton, 
have not succeeded in detecting an Fe-K line. 
Cold Fe in the interior of SN~1006 is known to exist based on ultraviolet 
absorption studies.
Blue and red-shifted Fe\emissiontype{II} absorption lines 
were detected in the spectra of background stars 
(Wu et al.~1993; Hamilton et al.~1997; Winkler et al.~2005), and 
interpreted as being due to unshocked Fe in the interior of the remnant. 
These results show that portions of the Fe-rich ejecta are 
still expanding freely, and have not yet been overrun by the reverse shock.
However the amount of Fe inferred in these studies ($<$ 0.16 M$_\odot$) is 
much less than the amount (0.6--0.8  M$_\odot$) predicted to be produced in
the thermonuclear disintegration of a $\sim$1.4 M$_\odot$ white dwarf.

In the early phases of supernova remnant evolution, even the shock-heated 
plasma is far from thermal equilibrium 
in terms of either the ionization or particle (electron and ion)
temperatures. 
The ionization age, a key diagnostic of
the non-equilibrium ionization (NEI) state, is defined as $n_et$, the
product of the electron density and the time since the gas was heated. 
Typically, $n_et$ is required to be $\geq 10^{12}$~cm$^{-3}$~s 
for full ionization equilibrium (Masai 1984). 
The thermal X-ray spectrum of SN~1006 suggests an ionization timescale of
$n_et\sim 2\times 10^9$~cm$^{-3}$~s (Vink et al.~2000), 
which is far from the full ionization equilibrium, 
and lower than nearly all other Galactic SNRs. 
The electron temperature at the northwest rim of the remnant
has been estimated to be $kT_e \sim 0.7$~keV (Long et al.~2003) 
or $\sim 1.5$~keV (Vink et al.~2003). 
Since these values are much lower than the ion temperature expected from 
the shock velocity of $\sim$2980~km~s$^{-1}$ (Ghavamian et al.~2002), 
non-equilibration of ion and electron temperatures is present in SN~1006 as well.

The extreme non-equilibrium state and high shock velocity are likely due to 
the low density of the ambient gas, 
because SN~1006 ($b=+14.6$) is further from the Galactic plane 
than other Galactic SNRs, such as Cas~A ($b=-2.1$) and
Tycho's SNR ($b=+1.4$). Although Kepler's SNR  ($b=6.8$) is nearly as far
above the plane as SN1006 (when their respective distances are included),
Kepler appears to be evolving into a dense circumstellar medium.
Therefore, the evolutionary state of SN~1006 may be the lowest of these
youngest Galactic SNRs, 
even though the real age is $\sim$3 times older than the others.

In this paper, using the type Ia SNR, SN~1006, we consider 
the X-ray spectra to investigate this very early phase of SNR evolution.
Suzaku (Mitsuda et al.~2007), with its high sensitivity for diffuse sources
and spectroscopic resolution in the $\sim$5--10~keV energy range, 
is well-suited for this study.

We assume the distance to SN~1006 to be 6.8$\times 10^{21}$~cm, 
following Winkler et al.~(2003).
The errors quoted in this paper represent the 90\% confidence level, 
unless otherwise stated.

\begin{figure}[tbh]
  \begin{center}
    \FigureFile(80mm,80mm){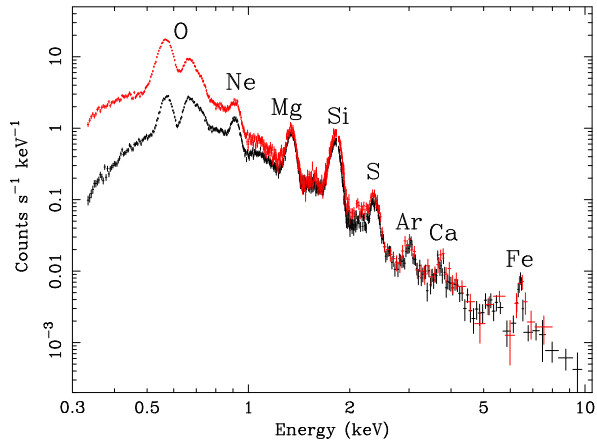}
  \end{center}
  \caption{Background-subtracted XIS spectra 
    extracted from the whole SE quadrant (SN~1006 SE).
    The black and red points represent the FI and BI spectra, respectively.}
  \label{fig:spectrum}
\end{figure}

\section{Observations and Data Reduction}
\label{sec:observation}

Four pointings were made of SN~1006 with Suzaku 
during the performance verification (PV) phase. 
Suzaku has one Hard X-ray Detector 
(HXD; Takahashi et al.~2007, Kokubun et al.~2007), 
and four X-ray Imaging Spectrometers 
(XIS; Koyama et al.~2007) 
each placed in the focal plane of an X-Ray Telescope
(XRT; Serlemitsos et al.~2007).
This paper reports on the imaging and spectral results obtained with the XIS. 
The XIS consist of three Front-Illuminated (FI) CCDs and 
one Back-Illuminated (BI) CCD. 
The advantages of the former are high detection efficiency and 
low background level in the energy band above $\sim$5~keV, 
while the latter has significantly superior sensitivity 
in the 0.2--1.0~keV band with moderate energy resolution. 
All four XRTs are co-aligned to image the same region of the sky.
The XIS were operated in the normal full-frame clocking mode 
with the editing mode of $3\times 3$ or $5\times 5$. 

We employed the cleaned revision 0.7 data, and used the HEADAS software 
version 6.0.4 and XSPEC version 11.3.2 for the data reduction and analysis. 
The X-ray data taken during low cut-off rigidity ($\leq$ 6~GV) were excluded 
so as to minimize any possible uncertainty in the non X-ray background (NXB). 
The total effective exposure times after these screening were $\sim$50~ks, 
for each of the pointings. 
The response matrix files (RMF) and ancillary response files (ARF) were made 
using xisrmfgen and xissimarfgen (Ishisaki et al.~2007) version 2006-10-17.

\section{Overall Structure}
\label{sec:overall}

\begin{figure*}[tbh]
  \begin{center}
    \FigureFile(80mm,80mm){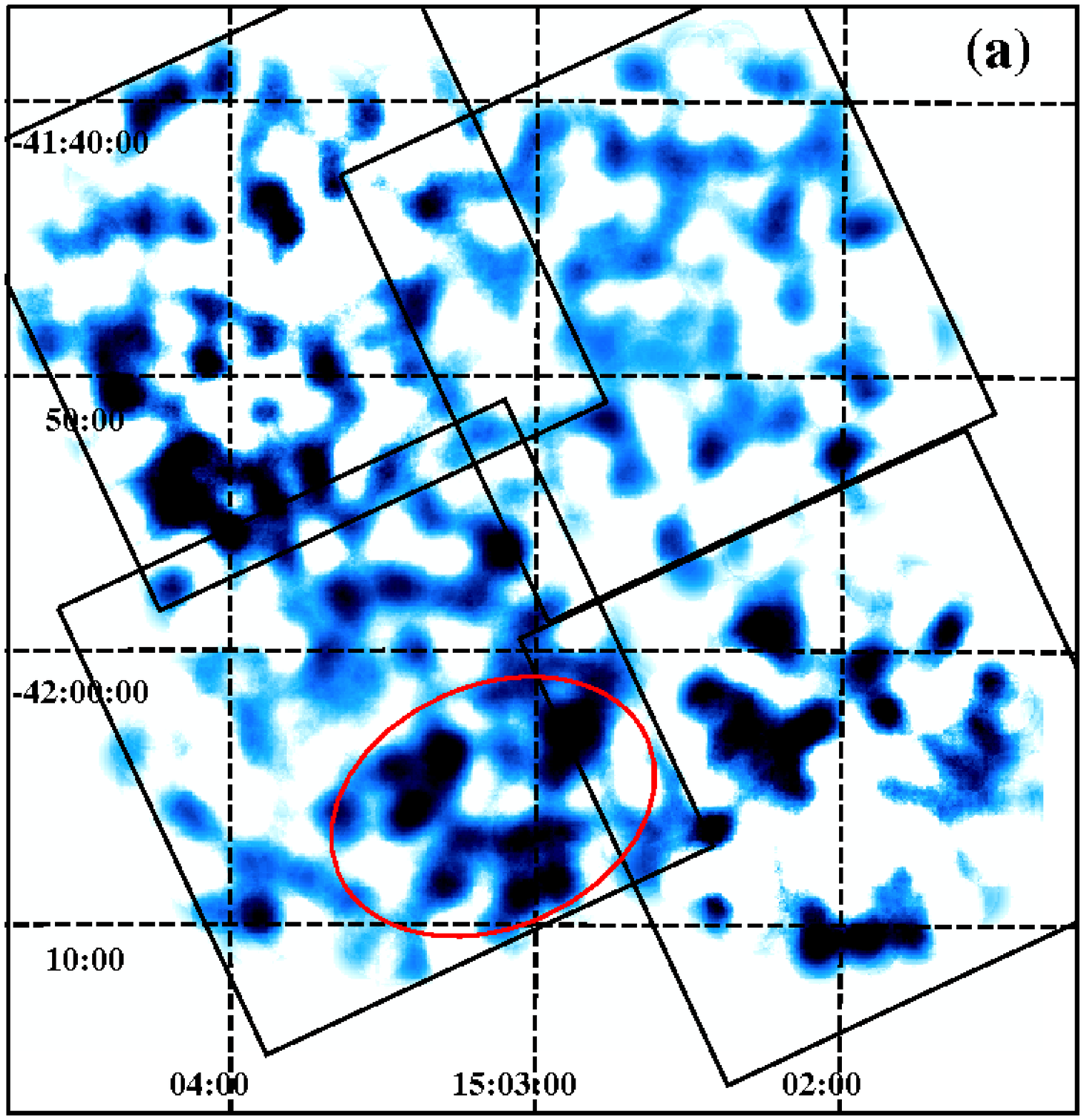}
    \FigureFile(80mm,80mm){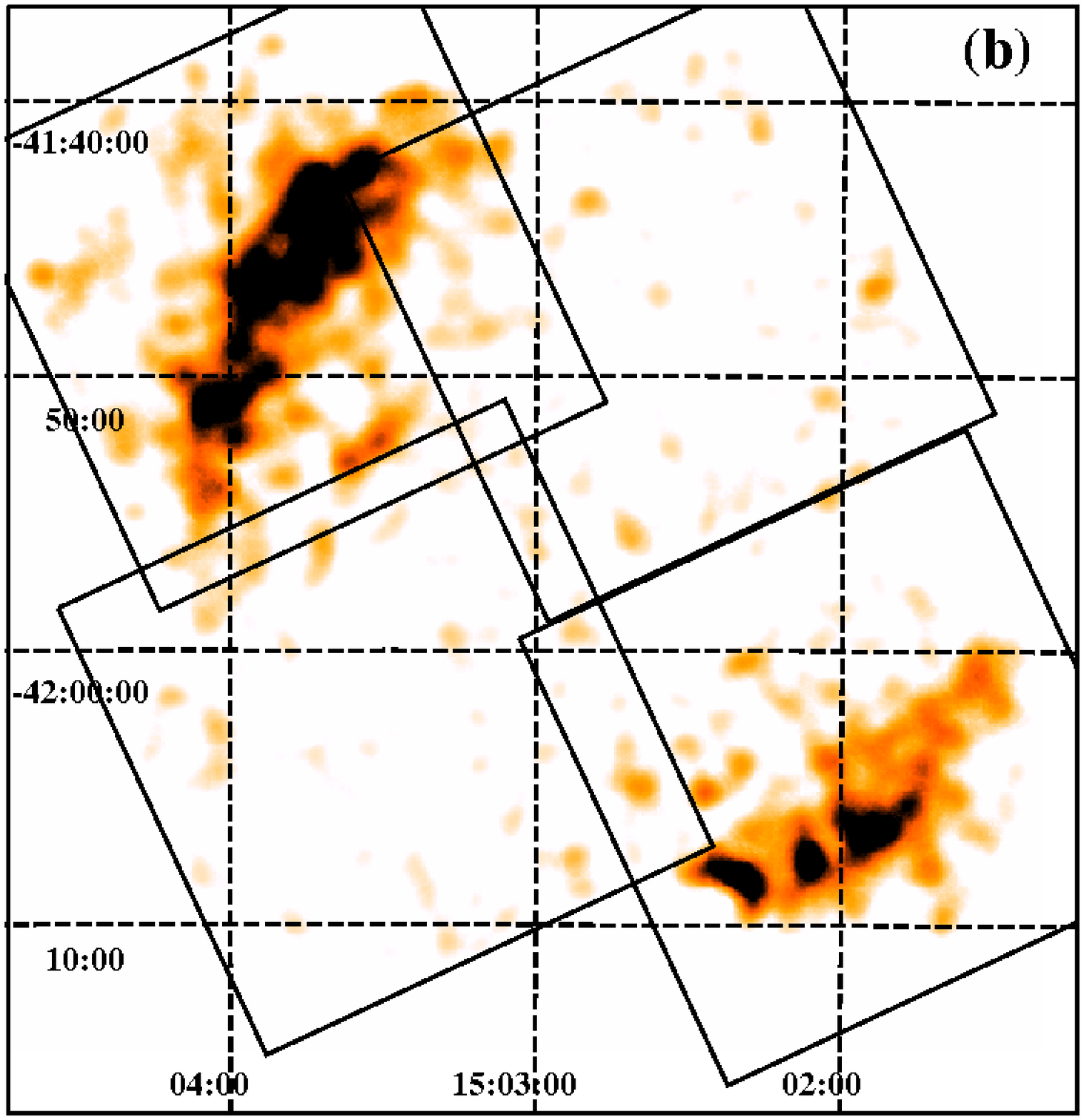}
  \end{center}
  \caption{XIS intensity map at the Fe-K$\alpha$ line 
    (a: 6.33--6.53~keV band), from which the continuum flux 
    at 6.1--6.3~keV band [shown in (b)] is subtracted.
    In both images, exposure and vignetting effects are corrected.
    The data from the three FIs are combined.
    Two corners of the calibration sources are removed.
    The black squares indicate each FOVs of the XIS.
    The red ellipse shows the region where we extracted the spectra 
    for a detailed analysis.}
  \label{fig:fe_image}
\end{figure*}

\subsection{Thin-Thermal Plasma Map}
\label{ssec:o_map}

The field of view (FOV) of the XIS is $\sim 18'\times 18'$, and hence 
a four-point mosaic can completely cover SN~1006 (about 30$'$ diameter). 
Since we intended to study the thermal plasma, 
we searched for the region of the remnant with the most prominent thermal 
emission. 
We hence made a mosaic image in the 
O\emissiontype{VII}~K$\alpha$ line band (0.55--0.6~keV), 
because this line is a major component of the thermal plasma in SN~1006. 
The result is shown in figure~\ref{fig:o_image}. 
We can see bright emission in this band in the northeast (NE) 
and southwest (SW) regions of the remnant. 
However, these are dominated by non-thermal emission 
(Koyama et al.~1995), from which it is difficult to extract thermal spectra.
Other regions are dominated by thermal emission. 
In particular, as shown in figure~\ref{fig:o_image}, 
the southeast quadrant (here SN~1006 SE) is brighter than the northwest (NW). 
Therefore, our study concentrated on this region. 
The spectral results of non-thermal rims (NE and SW) are reported by 
Bamba et al.~(2008), and those of the other regions (NW quadrant and 
the center of the remnant) will be reported in another paper in the future. 
The FOV of the XIS on SN~1006 SE is outlined by the black solid square 
in figure~\ref{fig:o_image}, 
which was observed  on 2006 January 30 (Obs.~ID=500016010).

\subsection{Spectrum of the SN~1006 SE Region}
\label{ssec:spectrum}

We extracted the spectrum of the entire SN~1006 SE region, 
excluding the two corners in the FOV that contain calibration source 
emission. 
For the background, we used the North Ecliptic Pole (NEP) data 
(Obs.~date $=$ 2006 February 10, Obs.~ID $=$ 500026010, 
Exp.~time $=$ $\sim$82~ks). 
Although we have been monitoring and correcting the increase of 
the Charge-Transfer Inefficiency (CTI) for the XIS (see Koyama et al.~2007), 
recovery of the energy resolution cannot be made. 
However, we can ignore the difference of the spectral resolution 
between the SN~1006 SE and NEP observations, 
because the latter observation was made only ten days after the former. 
To minimize the uncertainty due to the background subtraction, 
in particular that of NXB, we applied the same data-screening criteria to
both the SN~1006 SE and NEP observations, and took background  
spectra from the same detector coordinates as the source regions 
after excluding point-like sources detected in the NEP data. 
The background-subtracted spectra are shown in figure~\ref{fig:spectrum}. 
Since the data from the three FIs are nearly identical, 
we merged those individual spectrum to improve the photon statistics.

As shown in figure~\ref{fig:spectrum}, we found 
clear K-shell (K$\alpha$) lines from Ar, Ca, and Fe, for the first time. 
With a power-law plus Gaussian-line fit, we determined 
the line center energy of the Fe-K$\alpha$ to be $\sim$6.43~keV. 
This energy constrains the Fe ionization state to be approximately Ne-like.

\subsection{Iron Line Map}
\label{ssec:fe_map}

We show in figure~\ref{fig:fe_image}a
an image in a relatively narrow band (6.33--6.53~keV) 
that contains the Fe-K$\alpha$ line.  
This image was generated by subtracting the continuum flux 
at energies of 6.1--6.3~keV. 
(The image in this band is shown in figure~\ref{fig:fe_image}b.) 

We can see that the Fe-K$\alpha$ flux is enhanced at the southern part 
of the remnant (outlined in red with a ellipse), except for 
the NE and SW quadrants where the non-thermal emission is dominant. 
The mean surface brightness at 6.33--6.55~keV within the elliptical region is 
$8.5~(\pm 0.5)~\times 10^{-9}$~photons~cm$^{-2}$~s$^{-1}$~arcmin$^{-2}$, 
while that outside it (only in the SE and NW quadrants) is
$4.6~(\pm 0.3)~\times 10^{-9}$~photons~cm$^{-2}$~s$^{-1}$~arcmin$^{-2}$. 
In order to study the thin-thermal spectrum 
with the best S/N ratio for Fe-K line, 
we extracted the X-ray spectrum from within the elliptical region, 
excluding the corner of the calibration sources. 
The background subtraction was made in the same way 
as that of the full-field spectrum. 
The results are given in figure~\ref{fig:broad}. 
Hereafter, all detailed analyses are made using this spectrum.

\begin{table}[tbh]
  \caption{The center energies and widths of the emission lines.}
  \begin{center}
    \label{tab:line}
    \begin{tabular}{lcc}
        \hline 
        Line  &  Center energy$^{\ast}$~(eV) &  Width$^{\dagger}$~(eV) \\ 
        \hline
        Mg-K$\alpha$  & 1338 (1337--1340) & $<$ 5.4       \\
        Si-K$\alpha$  & 1815 (1813--1816) & 40 (38--42)  \\
        S-K$\alpha$   & 2361 (2355--2365) & 60 (54--65)  \\
        Ar-K$\alpha$  & 3010 (2991--3023) & $<$ 50        \\
        Ca-K$\alpha$  & 3692 (3668--3724) & $<$ 57        \\
        Fe-K$\alpha$  & 6430 (6409--6453) & $<$ 60        \\
        \hline
  \multicolumn{3}{l}{$^{\ast}$ Errors (statistical only) are given 
    in the parentheses}\\
  \multicolumn{3}{l}{\ (see text).} \\
  \multicolumn{3}{l}{$^{\dagger}$One standard deviation (1$\sigma$).} \\
  \end{tabular}
 \end{center}
\end{table}

\subsection{Energy and Width of the Emission Lines}
\label{ssec:line}

In order to study the line features, 
we fitted the spectra extracted from the elliptical region
with a phenomenological model; 
a power-law for the continuum and Gaussians for the emission lines. 
The best-fit central energies and widths 
for the emission lines are shown in table~\ref{tab:line}. 
Since the absolute energy calibration error is 
$\pm$0.2\% above 1~keV (Koyama et al.~2007), 
the relevant elements are uniquely identified 
from the best-fit central energy. 
These are also listed in table~\ref{tab:line}. 
Except for Mg, the central energies are significantly 
lower than that of the respective He-like K$\alpha$ lines. 
We also note that the widths of the Gaussians identified with 
Si-K$\alpha$ and S-K$\alpha$ are significantly broader than 
the instrumental energy resolution at those energies. 
Figure~\ref{fig:middle}a would help to compare the widths of the data 
with those of narrow lines.
In subsection~\ref{ssec:middle}, we discuss this matter in detail.

\section{Spectral Structure in the Narrow Bands}
\label{sec:narrow}

In order to study the plasma characteristics in relation to 
the various elements, we at first divided the spectrum 
into three representative energy bands: 
the 0.4--1.1~keV band for the O and Ne (light elements) lines, 
the 1.2--2.8~keV band for the Mg, Si, and S (medium elements) lines, 
and the 5--10~keV band for the Fe (heavy element) line. 
With the fitting to these individual band spectra, 
we constrained the plasmas including light, medium, 
and heavy elements separately.

\begin{figure}[tbh]
  \begin{center}
    \FigureFile(80mm,80mm){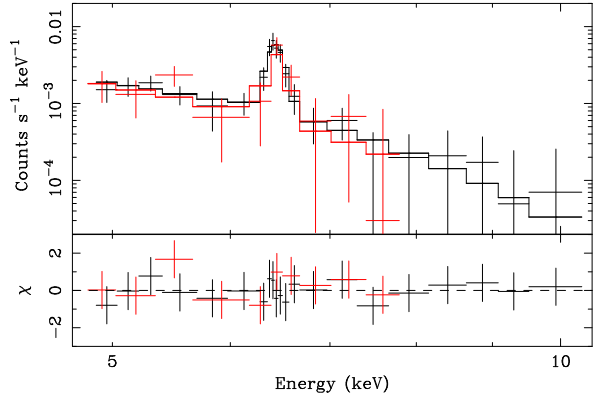}
  \end{center}
  \caption{XIS spectra in the 5--10~keV band with the VNEI model. 
    The black and red data points represent the FI and BI spectra, 
     respectively. 
    In the BI spectrum, the energy band above 8~keV is ignored.}
  \label{fig:high}
\end{figure}

\subsection{The 5--10~keV Band Spectra}
\label{ssec:high}

We fitted the 5--10~keV spectrum with a solar abundance 
(Anders and Grevese~1989) VNEI model.  Since the NEIvers~2.0
plasma code does not include K-shell emission lines 
for ions below the He-like state, we reverted to the NEIvers~1.1 code. 
If we fix the Fe abundance to be solar, 
then the model cannot reproduce the Fe-K$\alpha$ profile or flux, 
although the $\chi ^2$/d.o.f.~of 25/28 is acceptable. 
Allowing the Fe abundance to be free results in a greatly improved
best-fit $\chi^2$/d.o.f.~of 10/27.
The best-fit $kT_e$, $n_et$, and Fe abundance are 3.8~(1.7--26)~keV, 
6.1~(2.2--10)~$\times 10^9$~cm$^{-3}$~s, and 6.5~($>$1.9)~solar, respectively. 
The Fe abundance is significantly enhanced relative to solar, 
even if the continuum flux in this band is assumed 
to be composed of pure thermal emission. 
The best-fit spectrum is shown in figure~\ref{fig:high}.

\subsection{The 0.4--1.1~keV Band Spectra}
\label{ssec:low}

\begin{figure}[ptbh]
  \begin{center}
    \FigureFile(80mm,80mm){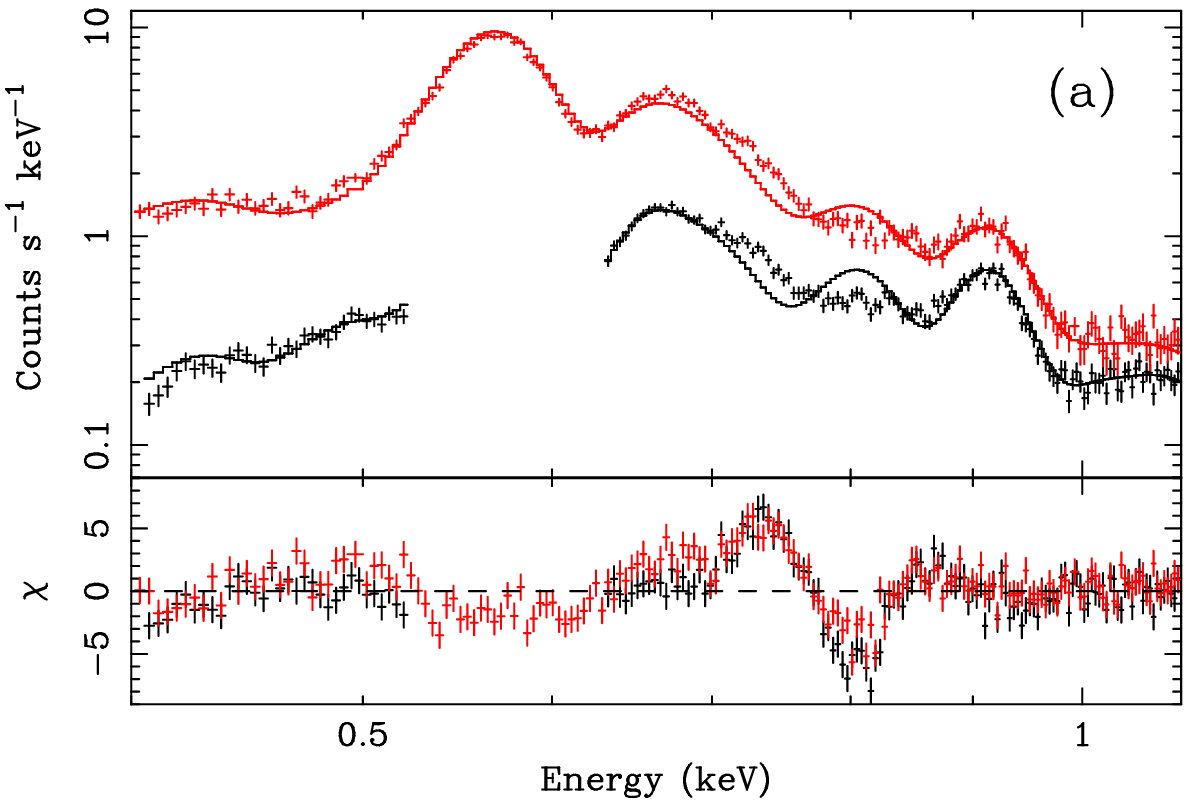}
    \FigureFile(80mm,80mm){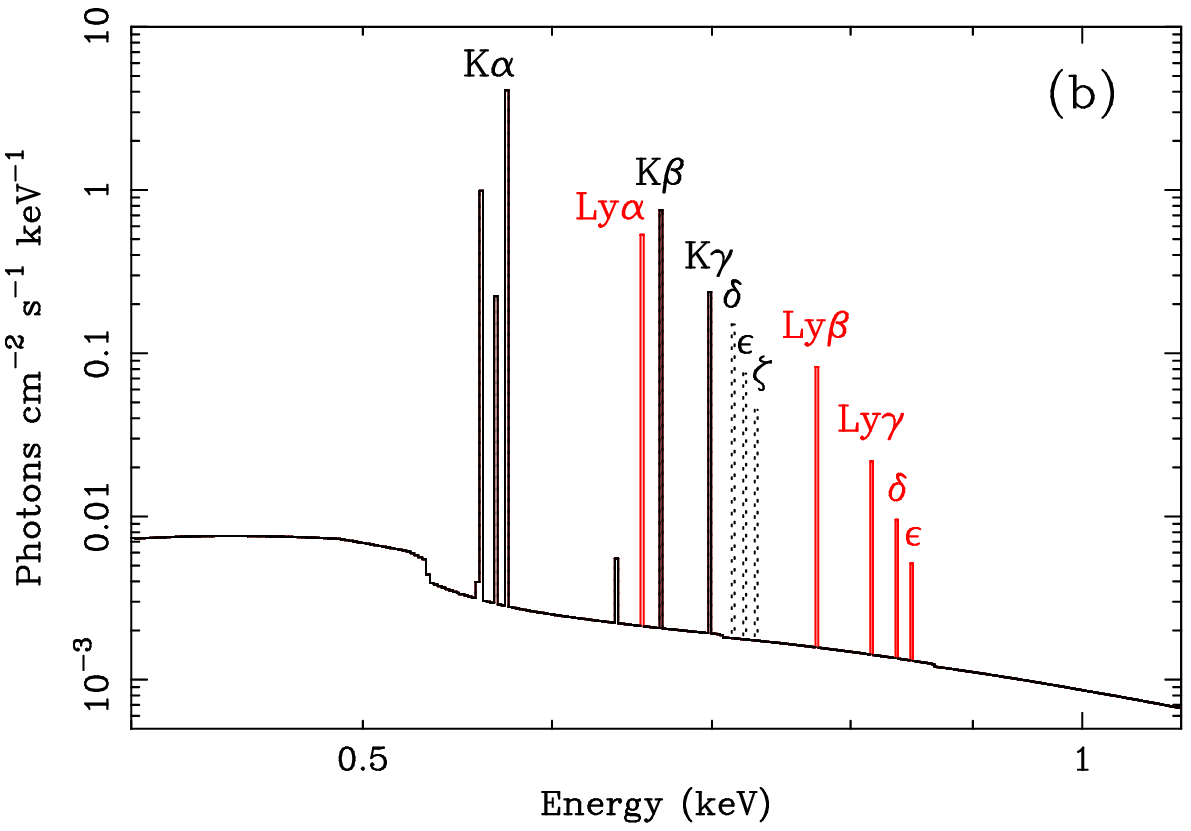}
    \FigureFile(80mm,80mm){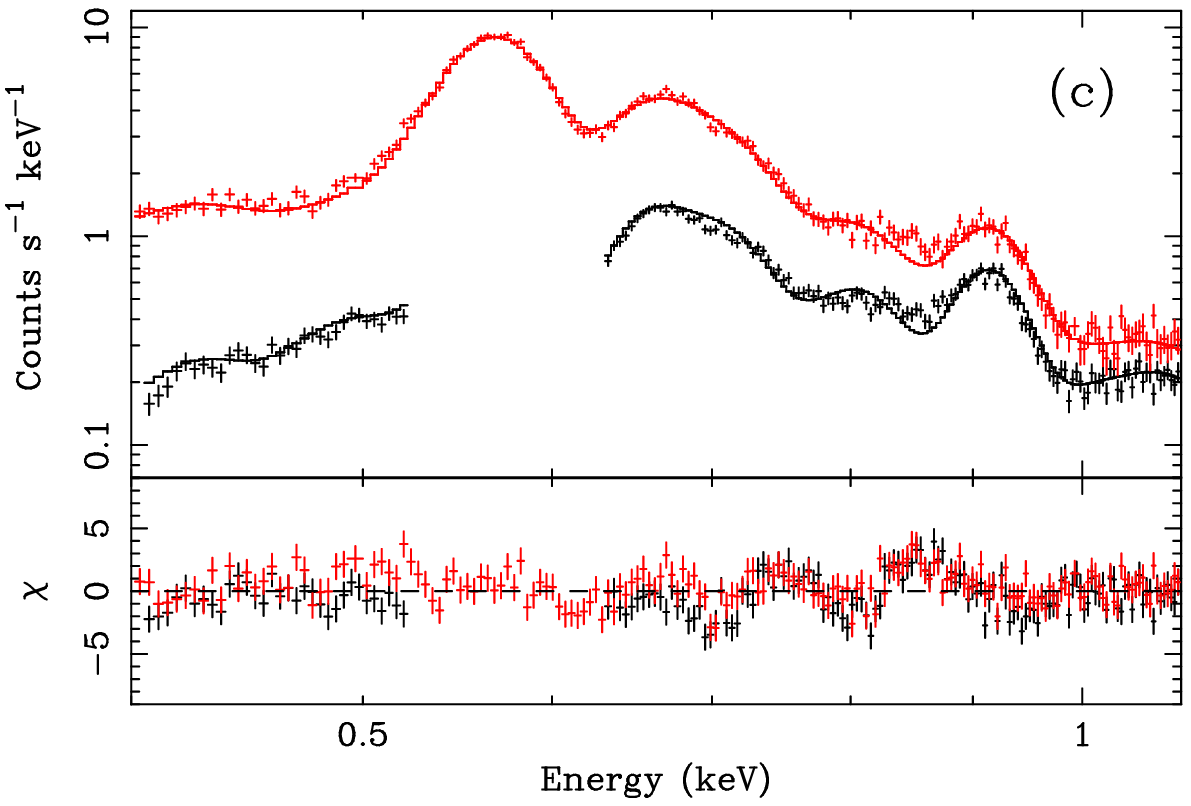}
  \end{center}
\caption{(a) XIS spectra in the 0.4--1.1~keV band 
  fitted with a VNEI model.
  The black and red data  points represent the FI and BI spectra, 
  respectively.  
  (b) Spectrum in photon space corresponding to the model 
  used in (a), but only showing O lines. It is binned every 2~eV.
  The black and red lines represent O\emissiontype{VII} series 
  and O\emissiontype{VII} series, respectively. 
  The dotted lines are additional transitions of K$\delta$ (714 eV), 
  K$\epsilon$ (723 eV), and K$\zeta$ (730 eV) of O\emissiontype{VII} 
  added as separate Gaussians (see text). 
  (c) Same spectra as (a), but for fits including Gaussians representing 
  K$\delta$ (714 eV), K$\epsilon$ (723 eV), and K$\zeta$ (730 eV) 
  of O\emissiontype{VII} (see text). 
  The residuals in the energy band 0.7--0.85~keV seen in (a) are largely
   removed.}
  \label{fig:low}
\end{figure}

Figure~\ref{fig:low} shows the 0.4--1.1~keV band spectra. 
This energy band includes K$\alpha$ lines of the light elements, 
with dominant emission from the oxygen K-shell line series. 
The spectra were fitted with a VNEI model, allowing the 
abundances of C, N, O, and Ne to be free parameters. 
Also, we let the abundances of Ca and Fe to be free, 
because the L-shell lines of these elements fall into 
this energy band (0.4--1.1~keV). 
Interstellar absorption was fixed to a hydrogen column density of
$6.8 \times 10^{20}$~cm$^{-2}$ (Dubner et al.~2002). 
In the initial fits, we found a significant inconsistency between the FI and 
BI spectra near the energy of the O-edge~(0.54~keV).
This may be due to incomplete calibration information for the 
contamination layer
on the optical blocking filter (OBF) of the XIS (Koyama et al.~2007). 
Since the calibration of the OBF contamination for the BI is 
more accurate than that for the FI, we decided to retain the BI
data across this energy band and
ignore the 0.5--0.63~keV band in the FI spectrum.
The results are shown in figure~\ref{fig:low}a, 
with the best-fit temperature and ionization parameter of 
$kT_e = $ 0.58~(0.56--0.59)~keV and 
$n_et = $ 6.7~(6.6--6.8)~$\times 10^9$~cm$^{-3}$~s.

In figure~\ref{fig:low}a, we find an apparent disagreement 
between the data and model in the  $\sim$0.7--0.85~keV energy band.
Usually, this energy band is dominated by the iron L-shell line transitions of 
$3s-2p$~($\sim$730~eV) and $3d-2p$~($\sim$830~eV) from Fe\emissiontype{XVII}. 
Although the fluxes of these two lines in NEI models are nearly equal over
a wide range of plasma conditions,
extrapolation to the extremely low ionization states that we see here
in SN~1006 is quite uncertain.  In particular, for the young Type Ia SNR 
E0509$-67.5$, Warren and Hughes (2004) found that the Fe L-shell emission 
was dominated by an emission from a line feature near 0.73 keV, which was 
not sufficiently strong in the spectral models and had to be included as
a separate Gaussian component.  Based on the strength of other Fe-L
emission lines and the weakness of O emission in E0509$-67.5$ the additional
line feature at 0.73 keV was confidently associated with Fe-L emission. 

Although we cannot eliminate the possibility that the residual comes
from faint Fe-L emission, here we consider a different explanation.
The best-fit spectral model of the oxygen lines are shown
in figure~\ref{fig:low}b, 
in which the black and red solid lines are K-shell transitions 
from O\emissiontype{VII} and O\emissiontype{VIII}, respectively. 
In  most astrophysical plasmas, 
the best-fit temperature of $\sim$0.6~keV gives weak K-shell lines 
of O\emissiontype{VII} compared to those of O\emissiontype{VIII}. 
On the other hand, in lower temperature plasmas 
(e.g., $kT_e \sim 0.1-0.2$~keV) 
where the O\emissiontype{VII} K$\alpha$ line is dominant, 
the line fluxes decrease rapidly along the K-shell transition series 
(K$\alpha, \beta, \gamma, \delta, \epsilon, \zeta$, etc.). 
Therefore, K-shell lines in the higher transitions can be 
safely ignored, and hence conventional NEI codes do not include 
O\emissiontype{VII} K-shell transition lines higher than K$\delta$. 
This is the reason that no oxygen line in the $\sim$0.7--0.85~keV band is 
present in the best-fit model (figure~\ref{fig:low}b). 
In SN~1006, however, the ionization timescale is low, 
while $kT_e$ is moderate, so the fluxes of higher level K-shell transitions 
from O\emissiontype{VII} may be relatively strong and cannot be ignored. 
To account for them we added Gaussians at 714~eV, 723~eV, and 730~eV 
to represent the O\emissiontype{VII} K$\delta$, K$\epsilon$, and K$\zeta$ 
lines, respectively. 
At a plasma temperature of 0.6~keV, 
the flux ratio of O\emissiontype{VIII}~Ly$\epsilon$ to Ly$\delta$ predicted
by the NEIvers~1.1 code is $\sim$0.5 (see figure~\ref{fig:low}b). 
In the absence of a detailed calculation we merely assumed that the flux 
ratio of the O\emissiontype{VII} lines follow the same pattern, namely
K$\epsilon$/K$\delta$ = K$\zeta$/K$\epsilon$ = 0.5.
The best-fit results with these additional lines,
see figure~\ref{fig:low}c, are a reasonably good fit.
The additional artificial lines are also 
shown as the dotted lines in figure~\ref{fig:low}b. 
The temperature and ionization parameter are almost the same as those 
obtained without the additional higher K-shell transitions of 
O\emissiontype{VII}. 

The best-fit intensity ratio of O\emissiontype{VII}~K$\beta$ to 
O\emissiontype{VIII}~Ly$\alpha$ is $\sim$1.4, 
which is consistent with the result of the XMM-Newton/RGS observation 
of the X-ray knot localized at the NW rim of SN~1006 (Vink et al.~2003; 
O\emissiontype{VII}~K$\beta$/O\emissiontype{VIII}~Ly$\alpha \sim$~1.6). 
This agreement may justify our approach to include the K-lines manually. 
Also, this suggests that the O\emissiontype{VII} K-shell line series are 
key spectral components for all over the SN~1006 plasma.

\begin{figure*}[tbh]
  \begin{center}
    \FigureFile(80mm,80mm){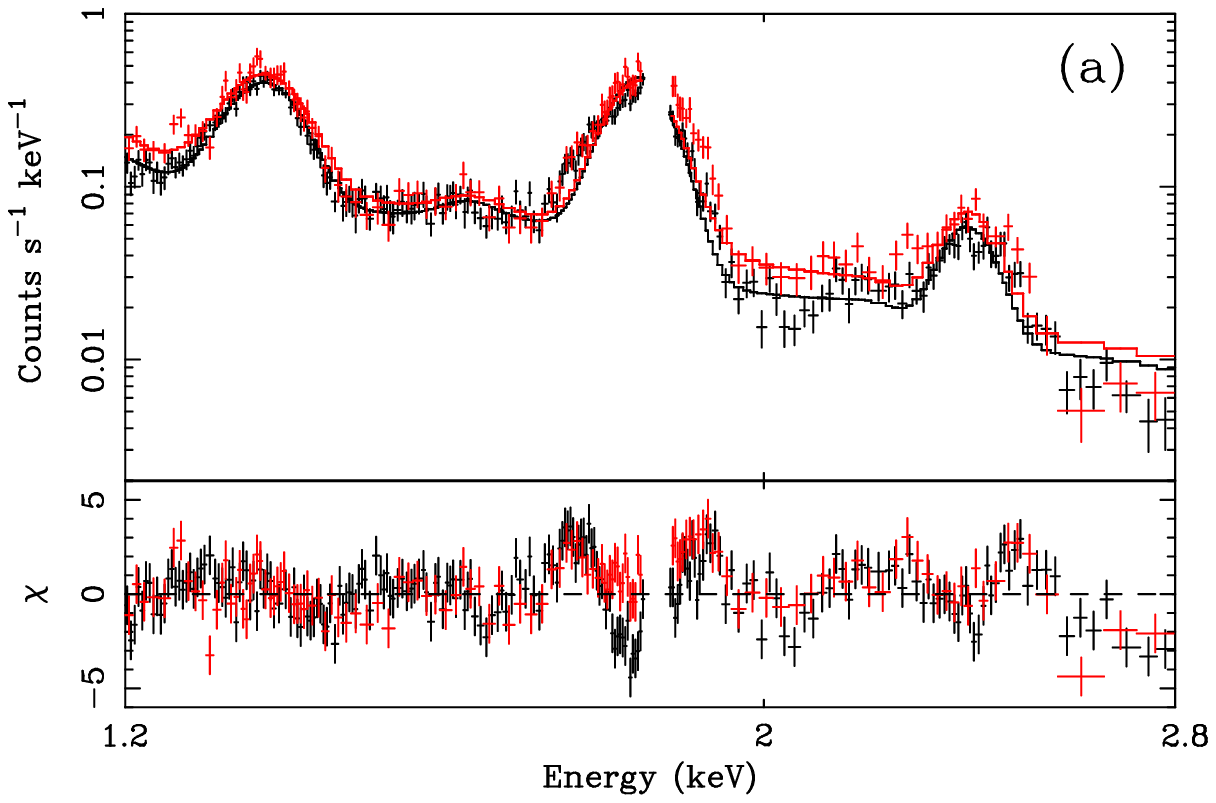}
    \FigureFile(80mm,80mm){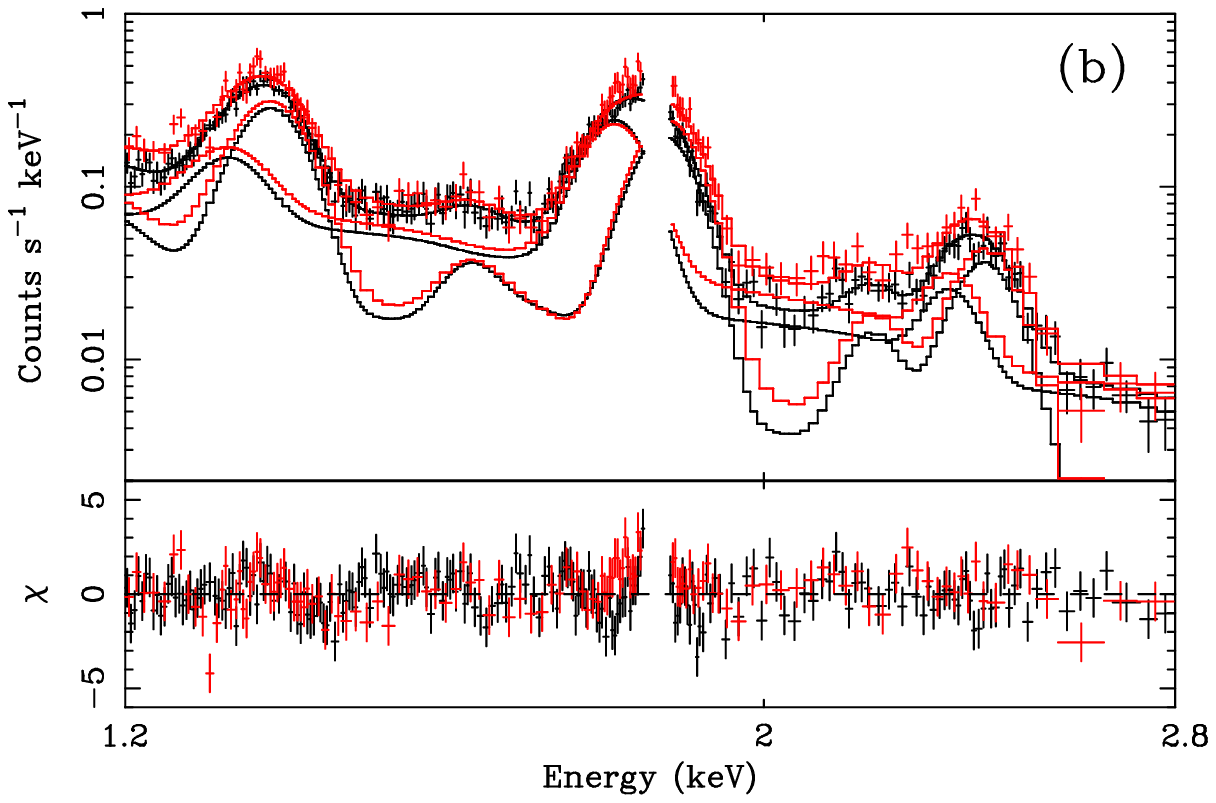}
  \end{center}
  \caption{(a) XIS spectra in the 1.2--2.8~keV band 
    fitted with the one-component VNEI model. 
    The black and red represent the FI and BI, respectively. \ 
    (b) Same to (a), but fitted with the two-component VNEI.}
  \label{fig:middle}
\end{figure*}

\subsection{The 1.2--2.8~keV Band Spectra}
\label{ssec:middle}

This band includes K-shell lines of medium-weight elements (Mg, Si, and S). 
As shown in table~\ref{tab:line}, we see can significant broadening of 
the Si and S K$\alpha$ lines. 
Therefore, the 1.2--2.8~keV band spectra cannot be reproduced with 
a single-component plasma model. 
In fact, we found significant disagreement between the data and model 
around the Si-K$\alpha$ and S-K$\alpha$ lines (figure~\ref{fig:middle}a) 
in the one-component VNEI fits. The lack of
data in the 1.83--1.85~keV band is due to current calibration errors of 
the XIS;  there is a small gap in energy ($\sim$10~eV) at the Si K-edge, 
which is not implemented in the current response function. 
The apparent line broadening of the Si K$\alpha$ line, however, 
is much larger ($\sim$40~eV) than this gap. 
Therefore, the disagreement is not due to a calibration error, but is real. 

If the intrinsic broadening of the Si-K$\alpha$ line
($\sim40$~eV; see table~\ref{tab:line}) is due to thermal Doppler broadening, 
the ion (silicon) temperature must be $\sim13$~MeV, 
which requires a shock velocity of $\sim 1.5\times 10^4$~km~s$^{-1}$. 
Ghavamian et al.~(2002) determined the shock velocity (from
measurements of the H$\alpha$ line width) to be $\sim$2900~km~s$^{-1}$
in the NW portion of SN~1006.  This is where the blast wave is
interacting with significant amounts of interstellar matter, 
compared to other parts of the rim. In fact, the average size of the
remnant ($\sim$\timeform{15.5'} radius) and the well-known age and
distance allow us to determine a mean expansion speed for SN~1006 to be
$\sim$$10^4$~km~s$^{-1}$.  The true current blast wave speed in SN~1006 
is almost surely bracketed by these two values. Furthermore, the reverse 
shock, which is likely to be the heating source for the Si that we see, 
typically moves into the ejecta at only a fraction of the blast wave speed.  
Therefore, we consider it unlikely that the line broadening we see
is due to thermal Doppler broadening.

\begin{table}[!b]
    \caption{Best-fit parameters and $\chi^2$ values of 
      the spectral fitting in the 1.2--2.8~keV band with
      the various models.}
  \begin{center}
    \label{tab:rejected}
    \begin{tabular}{lcccc}
      \hline 
      Model & $kT_{e1}$ & $kT_{e2}$  & $n_et$   & $\chi ^2$/d.o.f. \\
      ~ & (keV) & (keV) & (cm$^{-3}$~s) & ~ \\
      \hline
      1-VNEI    & 1.5  & --  & 4.5$\times 10^9$    & 848/351 \\
      2-VNEI    & 0.18 & 20  & 8.4$\times 10^9$    & 706/346 \\
      1-VPSHOCK & 1.6  & --  & 8.3$\times 10^9$    & 726/351 \\
      2-VPSHOCK & 0.36 & 12  & 1.0$\times 10^{10}$ & 597/346 \\
      \hline
    \end{tabular}
  \end{center}
\end{table}

Another possible explanation proposes that 
the emission consists of several thermal plasma components.
We fitted the spectra with 2-VNEI models, in which we allowed for 
different $kT_e$ values, but the same $n_et$ between components. 
However, no combination of parameters would fit the data, 
even if we let the abundances of Mg, Si, and S be free parameters. 
Not only 2-VNEI, but also 2-VPSHOCK models failed to fit the data when
the two components were forced to have the same ionization timescale.
All of the models that we tried, but rejected, are summarized 
in table~\ref{tab:rejected}. 
The addition of a power-law component to these models also 
did not help to improve the fittings.

\begin{table*}[tbh]
  \begin{center}
    \caption{Best-fit parameters of the spectral fitting 
      in the 1.2--2.8~keV band with 2-component VNEI model 
      with different $n_et$ values.}
    \label{tab:middle}
    \begin{tabular}{lcc}
      \hline 
      Parameter    & Component~1  & Component~2 \\
      \hline
      $N_{\rm H}$~(cm$^{-2}$) & 
               \multicolumn{2}{c}{6.8$\times 10^{20}$ (fixed)} \\
      $kT_e$~(keV)  & \multicolumn{2}{c}{1.1 (1.0--1.2)}  \\
      Mg        & 4.2 (3.5--5.2) & 6.1 (3.5--13)  \\
      Si        & 19 (15--24)    & 15 (13--24)    \\
      S         & 24 (19--34)    & 23 (17--31)    \\
      $n_et$~(cm$^{-3}$~s)  & 1.3~(1.1--1.7)~$\times 10^{10}$  &
                              7.9~(6.2--9.8)~$\times 10^8$     \\
      $n_{\rm H}n_eV$~(cm$^{-3}$) & 2.8~(2.3--3.4)~$\times 10^{55}$  & 
                              1.9~(1.6--2.3)~$\times 10^{56}$       \\
      \hline
      $\chi ^2$/d.o.f.  & \multicolumn{2}{c}{401/346 = 1.16}  \\
      \hline
    \end{tabular}
  \end{center}
\end{table*}

Next we tried  two-component VNEI models with 
the same electron temperature, but different $n_et$ values.
The best-fit reduced $\chi ^2$/d.o.f.~was greatly improved to 401/346 
(see table~\ref{tab:rejected}, for comparison). 
Models with different electron temperature and ionization timescale
(2-$kT_e$ and 2-$n_et$) gave no significant improvement 
of the reduced $\chi^2$. 
We thus conclude that a 2-component VNEI model with different 
$n_et$ values is necessary to fit the medium element plasma band. 
The best-fit parameters and spectra are given in table~\ref{tab:middle} 
and figure~\ref{fig:middle}b, respectively.

\begin{table*}[tbh]
    \caption{Plasma components determined from the narrow band spectra.}
   \begin{center}
   \label{tab:component}
    \begin{tabular}{lcccc}
      \hline 
      Band (keV) & Major elements & $kT_{e}$ (keV) & 
         $n_et$ (cm$^{-3}$~s) & Component$^{\ast}$ \\
      \hline
      0.4--1.1   & O, Ne      & 0.58 (0.56--0.59) & 
         6.7 (6.6--6.8)$\times 10^{9}$   & (3)  \\
      1.2--2.8   & Mg, Si, S  & 1.1 (1.0--1.2)    & 
         1.3 (1.1--1.7)$\times 10^{10}$  & (1)  \\
      ~          & ~          & 1.1 (1.0--1.2)    & 
	 7.9 (6.2--9.8)$\times 10^8$     & (2)  \\
      5.0--10    & Fe         & 3.8 (1.7--26)    & 
         6.1 (2.2--10)$\times 10^9$     & ~     \\
      \hline
      \multicolumn{5}{l}{$^{\ast}$The plasma component identifications 
                   described in text.} \\
    \end{tabular}
  \end{center}
\end{table*}

\section{Model Fit of the Full Band Spectra}
\label{sec:broad}

In section~\ref{sec:narrow}, we discuss how we separately derived 
spectral parameters for the three-energy bands that represent 
typical plasma conditions for the main light elements (O and Ne), 
medium elements (Mg, Si, and S), and heavy element (Fe). 
A summary is given in table~\ref{tab:component}. 
We used these parameters as the initial values to search for 
plasma parameters in the full energy band of 0.3--10~keV; 
the resulting overall best-fit parameters must be consistent with the
data over the entire XIS energy band. 

Based on table~\ref{tab:component}, we assume that the model that
describes the full spectral range must consist of, at least, three
plasma components: (1) high-$kT_e$ ($\sim$1~keV) with high-$n_et$
($\sim 10^{10}$~cm$^{-3}$~s), (2) similarly high-$kT_e$ with low-$n_et$
($\sim 10^9$~cm$^{-3}$~s), and (3) low-$kT_e$ ($\sim0.6$~keV) with
medium-$n_et$ ($\sim 7\times 10^9$~cm$^{-3}$~s).  First, we
consider whether or not the full band spectra can be well reproduced
with only these components.  We also consider a broader range of
$kT_e$ and $n_et$ values for the Fe-K emission.

As shown in table~\ref{tab:middle}, components (1) and (2) are
significantly enhanced in Mg, Si, and S (relative to solar).  
In the present spectral fits, we made the assumption that these components
are composed purely of metals without any admixture of hydrogen or
helium.  This assumption arose because hydrogen and helium emit
only continuum emission in the X-ray band.  Given the complexity of
our spectral model, it is just not possible with the XIS data to
obtain reliable estimates for the level of continuum emission from the
three plasma components separately. As a further assumption, 
we fixed the abundance of component (3) to be the solar value.
(Anticipating the results given below, we found that 
most of the continuum emission above 1 keV actually comes from a hard
power-law like spectral component.)
Operationally, we fixed the oxygen abundance in both the pure-metal 
components (i.e., nos.~1 and 2) to a large value ($1\times 10^4$), 
and fitted for the abundances of other elements relative to oxygen. 
The C and Ni abundances were assumed to be 
the same as Ne and Fe, respectively. 
The emission measure ($EM$) in the plasma code is defined (in XSPEC) as 
$n_{\rm H}n_eV$, even if the plasma is dominated by heavy elements. 
However, the oxygen density can be calculated as 
$n_{\rm O}$~=~($8.51\times 10^{-4}n_{\rm H}$)$\times 10^4$, where 
the numerical value is the solar abundance of oxygen 
from Anders and Grevesse~(1989).
Details of this method can be found in Vink et al.~(1996).

\begin{figure}[tbh]
  \begin{center}
    \FigureFile(80mm,80mm){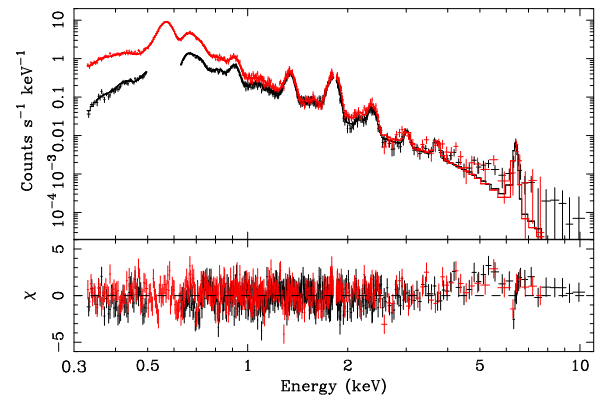}
  \end{center}
  \caption{XIS spectra in the 0.33--10~keV band 
    fitted with the three-component plasma model (see text).
    The black and red points represent the FI and BI data, respectively. 
    The model fails to describe the continuum emission above $\sim$4~keV.
  }
  \label{fig:no_pow}
\end{figure}

\begin{table*}[!p]
 \begin{center}
  \caption{Best-fit parameters of the broadband spectral fitting.}
  \label{tab:bestfit}
  \begin{tabular}{llll}
        \hline 
        Component  & \multicolumn{2}{c}{Parameter}        & Value   \\
        \hline
	Absorption & $N_{\rm H}$~(cm$^{-2}$) & ~ & 
				6.8$\times 10^{20}$ (fixed)         \\
	VNEI~1~(Ejecta~1) & $kT_e$~(keV)     & ~  & 1.2 (1.1--1.4)  \\
	~     & Abundance~($10^4$~solar)  & 
		      C  & 0.19$^{\ast}$ (0.08--0.30)    \\
	~     & ~   & N  & 0 (fixed)                        \\
	~     & ~   & O  & 1.0 (fixed)                      \\
	~     & ~   & Ne & 0.19$^{\ast}$ (0.08--0.30)    \\
	~     & ~   & Mg & 4.1 (3.7--4.7)                   \\
	~     & ~   & Si~~~ & 17 (15--20)                   \\
	~     & ~   & S  & 23 (20--27)                      \\
	~     & ~   & Ca & 11 (9.6--13)                     \\
	~     & ~   & Fe & 0.68$^{\dagger}$ (0.62--0.76)   \\
	~     & ~   & Ni & 0.68$^{\dagger}$ (0.62--0.76)   \\
	~     & $n_et$~(cm$^{-3}$~s)  & ~ & 1.4~(1.2--1.6)~$\times ~10^{10}$ \\
	~     & $n_{\rm H}n_eV$~(cm$^{-3}$)  & ~ &
				2.7~(2.4--3.0)~$\times ~10^{51}$  \\
	VNEI~2~(Ejecta~2) & $kT_e$~(keV)     & ~  & 1.9 (1.5--2.6)  \\
	~     & Abundance~($10^4$~solar)  
                     &C  & 0.33$^{\ddagger}$ (0.23--0.45)   \\
	~     & ~   & N  & 0 (fixed)                        \\
	~     & ~   & O  & 1.0 (fixed)                      \\
	~     & ~   & Ne & 0.33$^{\ddagger}$ (0.23--0.45)   \\
	~     & ~   & Mg & 3.0 (2.2--3.9)                   \\
	~     & ~   & Si~~~ & 10 (8.6--12)                  \\
	~     & ~   & S  & 12 (9.1--16)                     \\
	~     & ~   & Ca & 18$^{\S}$ (5.1--44)    \\
	~     & ~   & Fe & 18$^{\S}$ (5.1--44)    \\
	~     & ~   & Ni & 18$^{\S}$ (5.1--44)    \\
	~     & $n_et$~(cm$^{-3}$~s)  & ~ & 7.7~(6.7--9.2)~$\times ~10^8$ \\
	~     & $n_{\rm H}n_eV$~(cm$^{-3}$)       & ~ &
				1.8~(1.3--2.5)~$\times ~10^{52}$  \\
	NEI~(ISM)      & $kT_e$~(keV) & ~ & 0.51 (0.31--0.55)    \\
	~     & $n_et$~(cm$^{-3}$~s)  & ~ & 
				5.8~(5.7--6.1)~$\times ~10^9$     \\
	~     & $n_{\rm H}n_eV$~(cm$^{-3}$)  & ~ &
				1.1~(1.0--1.2)~$\times ~10^{56}$ \\
	Power-law & $\Gamma$  & ~         & 2.9 (2.8--3.0)       \\
	~     & \multicolumn{2}{l}{Norm$^{\|}$~(photons~cm$^{-2}$~s$^{-1}$)} &
				6.3~(5.5--7.2)~$\times ~10^{-4}$  \\
	\hline
	\multicolumn{4}{c}{Gaussian lines to complement the incomplete 
	  VNEI code (see text)}\\
	\hline
	Line & Center Energy(keV)(fixed) & & 
	Norm (photons~cm$^{-2}$s$^{-1}$)      \\
	\hline
	O\emissiontype{VII}-K$\delta$   & 0.714 & &
	6.4~(6.1--6.6)~$\times ~10^{-4}$  \\
	O\emissiontype{VII}-K$\epsilon$ & 0.723 & & 
	3.2~$\times 10^{-4~\#}$    \\        
	O\emissiontype{VII}-K$\zeta$    & 0.730 & & 
	1.6~$\times 10^{-4~\ast \ast}$   \\         
	Ar-K                            & 3.01  & & 
	5.7~(4.4--7.0)~$\times ~10^{-6}$  \\
	Ca-K			        & 3.69  & & 
	2.4~(1.3--3.5)~$\times ~10^{-6}$  \\
        \hline
	Gain   & Offset~(eV)   & ~  & --4.8            \\
	$\chi ^2$/d.o.f.  & ~  & ~  & 996/831 = 1.20   \\
	\hline 
   \multicolumn{4}{l}{$^{\ast}$, $^{\dagger}$, $^{\ddagger}$ 
		and $^{\S}$ represent linked parameters.} \\
   \multicolumn{4}{l}{$^{\|}$The differential flux at 1~keV.} \\
   \multicolumn{4}{l}{$^{\#}$Fixed to 50\% of the normalization of 
			 O\emissiontype{VII}-K$\delta$.} \\
   \multicolumn{4}{l}{$^{\ast \ast}$Fixed to 50\% of the normalization of 
			 O\emissiontype{VII}-K$\epsilon$.} \\
  \end{tabular}
 \end{center}
\end{table*}

\begin{figure*}[tbh]
  \begin{center}
    \FigureFile(140mm,140mm){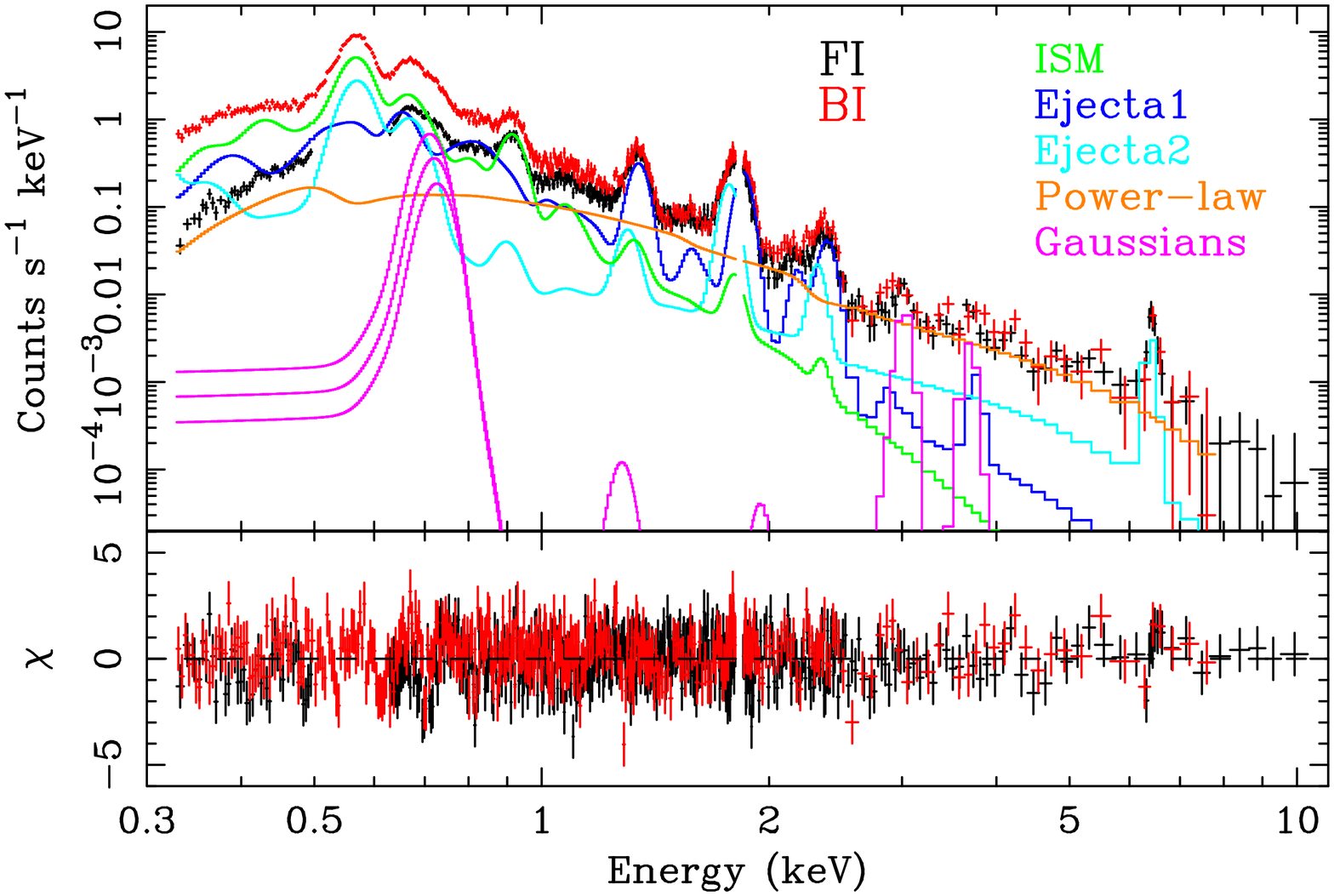}
  \end{center}
  \caption{XIS spectra in the 0.33-10~keV band
    which were extracted from the red elliptical region 
    shown in figure~\ref{fig:fe_image}.
    The black and red data points represent the FI and BI spectra, 
    respectively.
    The solid lines, colored green, blue, light blue, orange, and magenta 
    show the components of the ISM, Ejecta~1, Ejecta~2, Power-law, and 
    additional Gaussians of the best-fit model for the BI spectrum.
    Note that the Gaussian like structures (magenta) at 
    $\sim$1.2~keV and $\sim$1.9~keV are escape peaks of the added Gaussians 
    at 3.02~keV and 3.69~keV due to the XIS response.}
  \label{fig:broad}
\end{figure*}

Since the NEIvers~1.1 spectral model does not include 
K-shell emission lines from Ar and low ionization states of Ca, 
we added Gaussians at the energies of their  K$\alpha$ transitions:
3.02~keV (Ar) and 3.69~keV (Ca) (see table~\ref{tab:line}), 
For component (1) we additionally freely varied the Ca abundance, 
which effectively allowed for the Ca L-shell lines 
to contribute at low energies ($<$ 0.5 keV).
For component (2), we tied the Ca abundance to that of Fe, which was
allowed to vary freely.
We also added Gaussians for the upper-level oxygen transitions at 0.714, 
0.723, and 0.730~keV 
with the same reason as described in subsection~\ref{ssec:low}.

We fitted the FI data in the energy band from 0.33 to 10~keV 
(0.33 to 8~keV for the BI data).
We ignored the band below the C-edge at 0.28~keV, 
because of the large uncertainty in calibration 
due to contamination on the OBF 
(see Koyama et al.~2007). 
Energies of 0.5--0.63~keV (only for FI) and 1.83--1.85~keV were 
also ignored for the same reasons as noted in 
subsections \ref{ssec:low} and \ref{ssec:middle}. 
Since the absolute gain of the XIS has an uncertainty of 
$\pm$5~eV (Koyama et al.~2007), we allowed for a small offset in the
photon-energy to the pulse-height gain relationship.

With this model and assumptions, the best-fit 
$\chi ^2$/d.o.f.\ obtained was 1081/833. 
The fitted spectra are shown in figure~\ref{fig:no_pow}. 
All line features are well fitted, even Fe-K, suggesting that including
additional thermal components would not be justified.
However, there is a systematic data excess in the continuum at
energies $\gtrsim$~4~keV. We include a power-law component to model 
this, since we cannot reject the possibility that there is
a non-thermal component here that is similar to the  bright NE 
and SW rims  (albeit much less intense).
The fit with the additional power-law was significantly improved 
with $\chi ^2$/d.o.f.~=~996/831. 
The best-fit parameters are given in table~\ref{tab:bestfit}. 
As already noted, the gain may have an uncertainty of $\pm$5~eV. 
The fitting process requires a gain  offset of 4.8~eV 
for both FI and BI, within the allowable range. 
The best-fit model in the full energy band is shown in figure~\ref{fig:broad}.

\section{Discussion}
\label{sec:discussion}

In section~\ref{sec:broad}, we analyzed the full energy band spectrum 
extracted from the elliptical region shown in figure~\ref{fig:fe_image},
and found that it could be described adequately with a parameterized model
including three thermal plasmas in non-equilibrium ionization (VNEI~1,
VNEI~2, and NEI, in table~\ref{tab:bestfit}) and one power-law
component (Power-law, in table~\ref{tab:bestfit}).  Two of the 
plasmas, VNEI~1 and VNEI~2, were assumed to have non-solar abundances,
and the other, NEI, was assumed to have solar composition. 
In the following we discuss the implications of our spectral 
results, but the reader is cautioned that these results are subject 
to change if the assumptions we have made are changed.

\subsection{Origin of the Plasmas}
\label{ssec:origin}

The NEI model, assumed to have solar abundances, produces most of the
observed low-energy X-rays, particularly the K$\alpha$ lines from
O\emissiontype{VII}, O\emissiontype{VIII}, and Ne\emissiontype{IX}
(see figure~\ref{fig:broad}).  This component is rather uniformly
extended over the entire remnant, as can be seen in the O\emissiontype{VII}
line band map (figure~\ref{fig:o_image}).  We associate this component
with the swept-up ISM.

The other plasmas, VNEI~1 and VNEI~2, with their non-solar elemental 
abundance ratios, are plausibly of ejecta origin. 
Since VNEI~1 has a larger ionization parameter than the other plasmas, 
we suggest that this plasma was heated by reverse shock 
in the early stage of remnant evolution (here Ejecta~1). 
On the other hand, VNEI~2 has an extremely low ionization parameter, 
and hence should have been heated much more recently (here Ejecta~2). 
We have firmly detected iron K$\alpha$ emission for the first time. 
The low ionization state of Fe, plus its overabundance in the
Ejecta~2 component, is generally consistent with the Type Ia origin
of SN 1006.

\begin{figure*}[tbh]
  \begin{center}
    \FigureFile(80mm,80mm){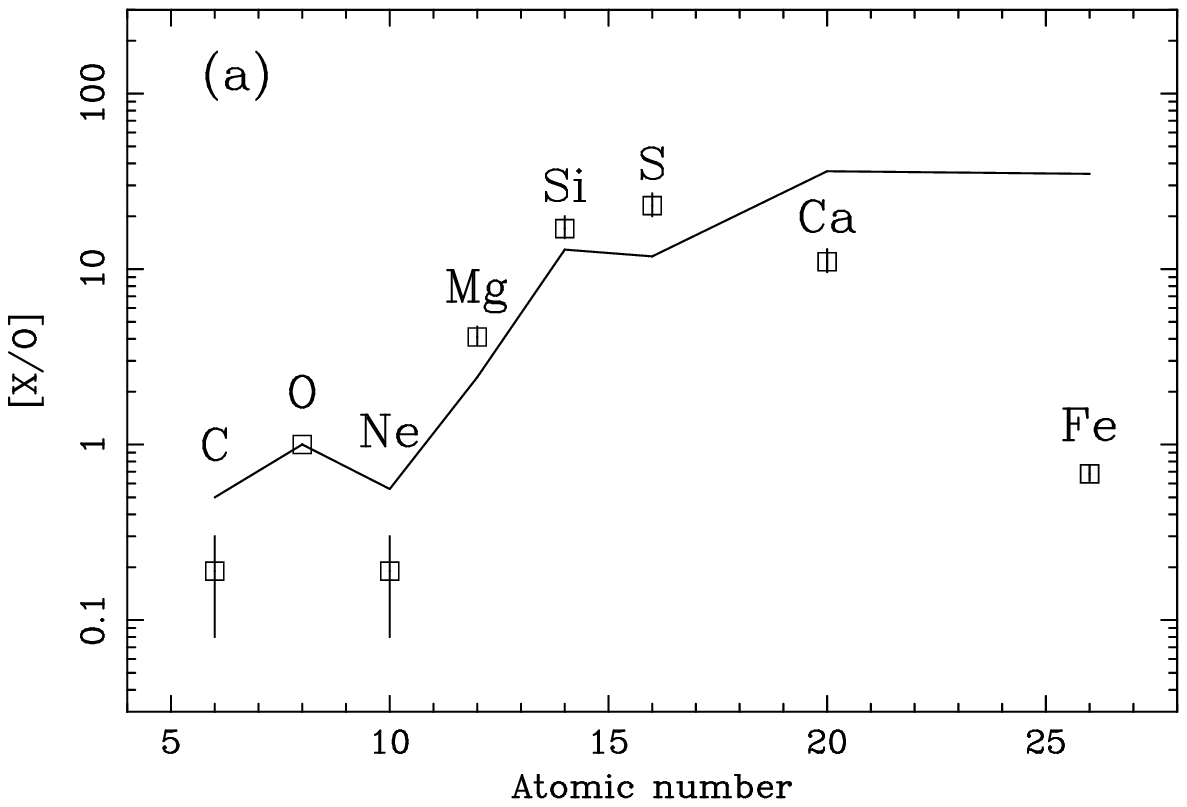}
    \FigureFile(80mm,80mm){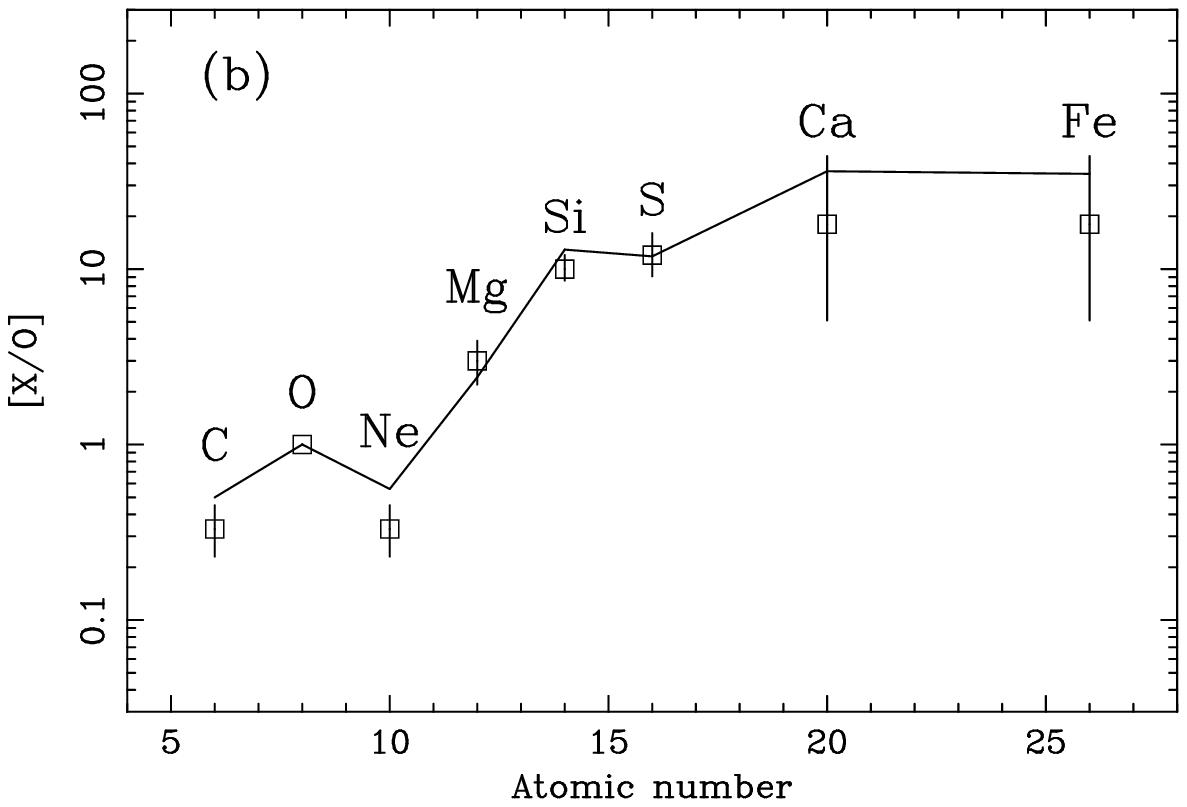}
  \end{center}
  \caption{Metal abundances as a function of atomic number 
    derived from the spectral fitting.
    The data points of (a) and (b) represent 
    those of  Ejecta~1 and Ejecta~2, respectively.
    The solid lines show the abundance relative to oxygen 
    calculated in the W7 model for a Type Ia supernova 
    by Nomoto et al.~(1984).}
  \label{fig:abundance}
\end{figure*}

\subsection{Relative Abundance in the Ejecta}
\label{ssec:abundance}

We compare our best-fit relative abundances with the predicted 
nucleosynthesis
yield of the widely-used W7 SN Ia model (Nomoto et al.~1984). 
For Ejecta~1, although the abundances of C, Ne, Mg, Si, and S 
relative to O are broadly consistent with the W7 model, 
Ca and Fe fall far below their predicted values, 
as shown in figure~\ref{fig:abundance}a. 
In the case of Ejecta~2, on the other hand, 
the heavy elements, albeit with large errors, are
broadly consistent with the abundance pattern from the W7 model, 
as shown in figure~\ref{fig:abundance}b. 
These results, along with the difference in the ionization timescale
between the components just mentioned, are consistent with a layed
composition of the ejecta with the higher-Z elements more concentrated
toward the center of SN~1006.

For this simple interpretation, one expects the
Fe line emission from the low-ionization timescale component
to be spatially located interior to the Mg and Si emission from
the high-ionization component, i.e., near the center of SN~1006.
As shown in figure~\ref{fig:fe_image}, though,
the Fe flux appears to peak close to the southeastern rim.
In summary, our two-component spectral model for the ejecta is a highly 
simplified view of what is surely a complex multi-$n_et$ and
multi-$kT_e$ structure that varies throughout the interior of the SNR.
A longer observation, particularly of the inner regions of SN~1006, 
will help to improve the statistical accuracy of the Fe-K line flux, 
and allow us to study this apparent discrepancy in more detail.

\subsection{ISM Density and Ion Temperature}
\label{ssec:ism}

The plasma parameters given in table~\ref{tab:bestfit} were derived from 
the southeast solid ellipse of figure~\ref{fig:fe_image}, which is 
$\pi \times \timeform{6.1'} \times \timeform{4.4'}$ = 84~arcmin$^2$ in area. 
This corresponds to 3.3 $\times ~10^{38}$~cm$^2$, 
from which we estimated the emission volume to be 
$V$ = (3.3 $\times ~10^{38}$)$^{3/2}$ = 6.0 $\times ~10^{57}$~cm$^3$. 
Therefore, the $EM$ of the heated ISM derived from the spectral fits
corresponds to an electron density of $n_e$ = 0.15~$f^{-0.5}$~cm$^{-3}$, 
where $f$ is the filling factor. 
Since the age of SN~1006 is $\sim$1000~yr, the ionization timescale can be 
roughly estimated as $n_et\sim 4.7\times 10^9$~cm$^{-3}$~s, 
which is almost consistent with the best-fit value of 
the ionization parameter of the NEI (ISM) component.
Such a low ionization age suggests that the temperatures of 
the ions and electrons may also be far from equilibrium. 
According to equation (3) of Laming~(2001), under the assumption that 
the initial ratio between $kT_e$ and $kT_{\rm H}$ just at the shock front is 
the ratio of the electron and proton masses, 
the proton temperature is estimated to be,
\begin{eqnarray}
  kT_{\rm H} = 7.8\left( \frac{n_et}{5.8\times 10^9~{\rm cm}^{-3}{\rm s}}
  \right)^{-1}\left( \frac{kT_e}{0.51~{\rm keV}}\right )^{5/2}~{\rm [keV]~.} 
  \nonumber
\end{eqnarray}
This result is consistent with that of the H$\alpha$ observation of 
$T_e/T_{\rm H} \leq 0.07$ (Ghavamian et al.~2002). 
The extreme non-equilibrium state of the plasma is due to 
the low density of the ambient medium. 
According to equations (2) and (5) of Ferri{\`e}re~(2001), 
the density of H\emissiontype{I} and H\emissiontype{II} 
at the $Z$-height of SN~1006 ($Z$ = 550~pc) is $\sim$0.03~cm$^{-3}$. 
This is consistent with our estimate of the ambient density of
$n_{\rm H}/4 \sim 0.03$~cm$^{-3}$.

\subsection{The Power-Law Component}
\label{ssec:power-law}

The low density of the ambient medium will allow the  
velocity of the shock front to remain high for a relatively long time. 
This may be one reason why this remnant shows such efficient particle 
acceleration up to very high energies ($\sim10^{14}$~eV), 
observed as power-law (synchrotron) emission 
from the NE and SW rims (Koyama et al.~1995). 
The photon index from these rims is $\Gamma \sim$ 2.7--2.9. 
Now here, away from the bright rims, we also detect a power-law component 
in the hard energy band (the Power-law, in table~\ref{tab:bestfit}) with
a best-fit index of $\Gamma \sim$2.8--3.0, 
which is consistent with the NE and SW bright rim emission. 
Since the power-law emissions from the rims are most likely to be 
synchrotron X-rays, the power-law component newly found from 
the ellipse in figure~\ref{fig:fe_image} is also likely to be synchrotron
X-ray emission. 
The surface brightness is estimated to be 
$\sim 6\times 10^{-15}$~ergs~cm$^{-2}$~s$^{-1}$~arcmin$^{-2}$ 
in the 2--10~keV band.
This value is $\sim$ 50-times lower than the peak emission 
at the NE and SW rims. 
If similar to the NE rim, the power-law component should show
thin filamentary structures (Bamba et al.~2003). 
However, XMM-Newton (Rothenflug et al.\ 2004) and Chandra 
(Hughes et al., in preparation) observations found little evidence 
for filamentary X-ray rims along the SE rim.

\section{Summary}
\label{sec:summary}

We have analyzed high-quality spectra obtained with Suzaku of a region 
in the southern part of SN~1006, 
selected because it is bright in Fe-K$\alpha$.
The results and interpretations are summarized as follows:

\begin{enumerate}

\item  The spectrum can be described by a model with at least three NEI
  thermal plasmas 
  and one power-law component.

\item  The fits yield different ionization parameters of 
  $n_et \sim 8\times 10^8$~cm$^{-3}$~s (low), 
  $\sim 6\times 10^9$~cm$^{-3}$~s (medium), 
  $n_et \sim 10^{10}$~cm$^{-3}$~s (high). 

\item  The low-$n_et$ plasma are highly overabundant in heavy elements, 
  in which we found K$\alpha$ lines from Fe, for the first time. 
  The abundance pattern is consistent with that of type Ia SN ejecta. 

\item  The abundance of the medium-$n_et$ plasma is assumed to be solar, 
  and we associate this component with the shocked ISM. Although oxygen 
  is not overabundant, 
  K$\alpha$ lines of O\emissiontype{VII} and O\emissiontype{VIII} 
  from this plasma appear to dominate the thermal emission from SN~1006. 
  (However the complexity of our spectral model makes this
  claim less than definitive. We cannot exclude the possibility that
  the oxygen emission from this ISM component actually comes from yet
  another ejecta component.  Spatially resolved spectroscopy would help
  to resolve this concern.)

\item  The high-$n_et$ plasma is overabundant in medium elements, 
  like Mg, Si, and S, but heavy element like Fe. 
  Our interpretation is that this plasma also has an ejecta origin, 
  and the composition is dominated by lower atomic number species.

\item  Temperature equilibrium between ions and electrons
  has not yet been achieved. 
  The proton temperature of the shocked ISM is $\sim$15-times higher 
  than the electron temperature. 

\item  The spectrum from this region contains a power-law component. 
  The photon index, $\Gamma \sim$2.9, is similar to that of the 
  northeast and southwest bright rims, suggesting that 
  this component is also of synchrotron origin.

\end{enumerate}

\bigskip

The authors thank all member of the Suzaku team, 
especially Y. Hyodo and H. Matsumoto.
H.Y., S.K., H.N., and A.B. are supported by JSPS Research Fellowship 
for Young Scientists.



\begin{thebibliography}{}

\bibitem[Anders \& Grevesse(1989)]{1989GeCoA..53..197A} Anders, E., \& 
Grevesse, N.\ 1989, \gca, 53, 197 

\bibitem[Bamba et al.(2003)]{2003ApJ...589..827B} Bamba, A., Yamazaki, R., 
Ueno, M., \& Koyama, K.\ 2003, \apj, 589, 827 

\bibitem[Bamba et al.(2008)] -Bamba, A., et al.~2007, 
\pasj, 60 (\# 3179) 

\bibitem[Decourchelle et al.(2001)]{2001A&A...365L.218D} Decourchelle, A., 
et al.\ 2001, \aap, 365, L218 

\bibitem[Dubner et al.(2002)]{2002A&A...387.1047D} Dubner, G.~M., Giacani, 
E.~B., Goss, W.~M., Green, A.~J., \& Nyman, L.-{\AA}.\ 2002, \aap, 387, 
1047 

\bibitem[Ferri{\`e}re(2001)]{2001RvMP...73.1031F} Ferri{\`e}re, K.~M.\ 
2001, Reviews of Modern Physics, 73, 1031 

\bibitem[Ghavamian et al.(2002)]{2002ApJ...572..888G} Ghavamian, P., 
Winkler, P.~F., Raymond, J.~C., \& Long, K.~S.\ 2002, \apj, 572, 888 

\bibitem[Hamilton et al.(1997)]{1997ApJ...481..838H} Hamilton, A.~J.~S., 
Fesen, R.~A., Wu, C.-C., Crenshaw, D.~M., \& Sarazin, C.~L.\ 1997, \apj, 
481, 838

\bibitem[Hwang et al.(1998)]{1998ApJ...497..833H} Hwang, U., Hughes, J.~P., 
\& Petre, R.\ 1998, \apj, 497, 833 


\bibitem[Ishisaki et al.(2007)]{2007PASJ...59S.113I} Ishisaki, Y., et al.\ 
2007, \pasj, 59, 113 


\bibitem[Iwamoto et al.(1999)]{1999ApJS..125..439I} Iwamoto, K., Brachwitz, 
F., Nomoto, K., Kishimoto, N., Umeda, H., Hix, W.~R., \& Thielemann, F.-K.\ 
1999, \apjs, 125, 439 

\bibitem[Kokubun et al.(2007)]{2007PASJ...59S..53K} Kokubun, M., et al.\ 
2007, \pasj, 59, 53 

\bibitem[Koyama et al.(1995)]{1995Natur.378..255K} Koyama, K., Petre, R., 
Gotthelf, E.~V., Hwang, U., Matsuura, M., Ozaki, M., \& Holt, S.~S.\ 1995, 
\nat, 378, 255 

\bibitem[Koyama et al.(2007)]{2007PASJ...59S..23K} Koyama, K., et al.\ 
2007, \pasj, 59, 23 

\bibitem[Laming(2001)]{2001ApJ...546.1149L} Laming, J.~M.\ 2001, \apj, 546, 
1149 

\bibitem[Long et al.(2003)]{2003ApJ...586.1162L} Long, K.~S., Reynolds, 
S.~P., Raymond, J.~C., Winkler, P.~F., Dyer, K.~K., \& Petre, R.\ 2003, 
\apj, 586, 1162 

\bibitem[Masai(1984)]{1984Ap&SS..98..367M} Masai, K.\ 1984, \apss, 98, 367 

\bibitem[Mitsuda et al.(2007)]{2007PASJ...59S...1M} Mitsuda, K., et al.\ 
2007, \pasj, 59, 1 

\bibitem[Nomoto et al.(1984)]{1984ApJ...286..644N} Nomoto, K., Thielemann, 
F.-K., \& Yokoi, K.\ 1984, \apj, 286, 644 

\bibitem[Rothenflug et al.(2004)]{2004A&A...425..121R} Rothenflug, R., 
Ballet, J., Dubner, G., Giacani, E., Decourchelle, A., \& Ferrando, P.\ 
2004, \aap, 425, 121 

\bibitem[Schaefer(1996)]{1996ApJ...459..438S} Schaefer, B.~E.\ 1996, \apj, 
459, 438 

\bibitem[Serlemitsos et al.(2007)]{2007PASJ...59S...9S} Serlemitsos, P.~J., 
et al.\ 2007, \pasj, 59, 9 

\bibitem[Takahashi et al.(2007)]{2007PASJ...59S..35T} Takahashi, T., et 
al.\ 2007, \pasj, 59, 35 

\bibitem[Vink et al.(1996)]{1996A&A...307L..41V} Vink, J., Kaastra, J.~S., 
\& Bleeker, J.~A.~M.\ 1996, \aap, 307, L41 

\bibitem[Vink et al.(2000)]{2000A&A...354..931V} Vink, J., Kaastra, J.~S., 
Bleeker, J.~A.~M., \& Preite-Martinez, A.\ 2000, \aap, 354, 931 

\bibitem[Vink et al.(2003)]{2003ApJ...587L..31V} Vink, J., Laming, J.~M., 
Gu, M.~F., Rasmussen, A., \& Kaastra, J.~S.\ 2003, \apj, 587, L31 

\bibitem[Wu et al.(1993)]{1993ApJ...416..247W} Wu, C.-C., Crenshaw, D.~M., 
Fesen, R.~A., Hamilton, A.~J.~S., \& Sarazin, C.~L.\ 1993, \apj, 416, 247 

\bibitem[Warren \& Hughes(2004)]{2004ApJ...608..261W} Warren, J.~S., \& 
Hughes, J.~P.\ 2004, \apj, 608, 261 

\bibitem[Winkler et al.(2003)]{2003ApJ...585..324W} Winkler, P.~F., Gupta, 
G., \& Long, K.~S.\ 2003, \apj, 585, 324 

\bibitem[Winkler et al.(2005)]{2005ApJ...624..189W} Winkler, P.~F., Long, 
K.~S., Hamilton, A.~J.~S., \& Fesen, R.~A.\ 2005, \apj, 624, 189 

\end{thebibliography}
\end{document}